\documentclass[jgrga]{agutex}
\usepackage{graphicx}
\authorrunninghead{TATJANA ZIVKOVIC, KRISTOFFER RYPDAL}

\titlerunninghead{EL-NINO DYNAMICS}

\begin{document}

\title{ENSO dynamics: low-dimensional-chaotic or stochastic?}

\authors{T.  \v{Z}ivkovi\'{c},\altaffilmark{1} and
K. Rypdal, \altaffilmark{2}}

\authoraddr{T.  \v{Z}ivkovi\'{c}, 
(tatjana.zivkovic@irfu.se)}

\altaffiltext{2}{Department of Physics and Technology,
University of Troms{\o}, Norway.}
\altaffiltext{1}{Swedish Institute of Space Physics, University of Uppsala, Sweden}

%
%


\begin{abstract}
We apply a test for low-dimensional, deterministic dynamics to  the Ni\~{n}o 3 time series for the El Ni\~{n}o Southern Oscillation (ENSO). The test is negative, indicating that  the dynamics is high-dimensional/stochastic. However, application of stochastic forcing to a time-delay equation for equatorial-wave dynamics can reproduce this stochastic dynamics  and other important aspects of ENSO. Without such stochastic forcing this model yields low-dimensional, deterministic dynamics, hence these results emphasize the importance of  the stochastic nature of the atmosphere-ocean interaction in low-dimensional models of ENSO.
 \end{abstract}

\begin{article}
\section{Introduction}
Prediction is the ultimate goal of meteorology and climatology and the issue of predictability is crucial. Prediction in these disciplines are mostly probabilistic, but there may be different rationales for a probabilistic description. These rationales are intimately linked to the various meanings of the concepts of determinism, deterministic chaos, and stochasticity. The  fundamental laws of classical physics are deterministic, since the future and past of the state of such a system is uniquely given by the state at a given time. The majority of models of weather and climate are deterministic in this sense.  After the discovery of deterministic chaos, it has been realized that the deterministic evolution may be sensitive to the initial conditions, in the sense that small perturbations in the initial conditions grow exponentially with time. Predictability in such systems is limited by this exponential growth rate, given by the largest positive Lyapunov exponent of the system. From a practical viewpoint the most important  aspect of this insight is that predictability can be limited in this way, and may require a probabilistic description, not only in systems with a large number of degrees of freedom, but also in in simple, low-dimensional, nonlinear systems. 

Before the advent of chaos theory, unpredictability was considered to be a practical consequence of our inability to specify and solve the evolution equations for  the microscopic state of  high-dimensional systems. However, these days  high-speed computing allows us, not only to solve numerically low-dimensional, nonlinear problems, but also high-dimensional general circulation models of the climate system. Nevertheless, the problem of sensitivity to initial condition seems to persist, at least when it comes to weather prediction and climate variability on interannual to decadal time scales. The fluctuations of macroscopic variables around some time-averaged or ensemble-averaged state could be described in a probabilistic manner by representing them as stochastic processes or alternatively as  low-dimensional chaotic processes.
Thus, when dealing with data in climatology, either from observation or from large-scale simulations, the climate dynamicist will have to ask the question of whether prediction for the system/phenomenon of interest  is better  served  by a low-dimensional chaotic model of climate variability, or by a high-dimensional (stochastic) model. 

The answer to this question depends on whether the system/phenomenon represents a self-organization of the dynamics into an effective small number of degrees of freedom. One  approach is of course to reduce the high-dimensional model to a low-dimensional one via a series of approximations and simplifications, i.e., via theoretical model reduction. In many cases, this is very  demanding and many different reductions are possible, and it is hard to know whether a reduced model is what Nature abides to. Hence, there is a demand for methods by which it is possible to decide from the observation data whether the system dynamics can be uniquely projected onto an attractor in a low-dimensional phase space. Such methods exist, based on Taken's time-delay embedding theorem \citep{T81}, and from 1980 onwards there is a growing literature on reconstruction of chaotic attractors from time series and computation of attractor dimension and largest Lyapunov exponent \citep{DA}. The idea is to assume that the system state vector evolves according to the system of first-order ordinary differential equations
describing the trajectory ${\bf z}(t)$ on the $d$-dimensional attractor in a $p>d$-dimensional phase space. If the system is autonomous and 
 the attractor of the trajectory has dimension $d$, Takens'  time-delay method  \citep{T81} can be used to construct an $m>2d$-dimensional embedding space on which the attractor can be mapped continuously and one-to-one. In practice this method works only if the attractor dimension $d$ is reasonably low. Dynamical systems with a large number of independent or weakly dependent  degrees of freedom can only be described either by large-scale numerical simulation or by stochastic methods. For such systems the phase-space attractor is also high-dimensional and cannot be mapped one-to-one onto a low-dimensional  time-delay  embedding space. The computation of attractor dimension then typically fails to converge when embedding dimension $m$ is increased, but such convergence can be difficult to detect if the time series is short. \cite{KG92,KG93} devised a direct test for the existence of low-dimensional deterministic dynamics which is useful for short time series. This is the kind of test that will be employed in this paper to the instrumental time series for the El-Ni\~{n}o Southern Oscillation (ENSO).
 
 The remainder of the paper is structured as follows: In section (2) we briefly describe a time-delay equation for equatorial wave dynamics and ENSO, and in section (3) and (4) we review the time-delay  phase space reconstruction technique and the test for determinism. In section (5) these techniques are employed to the Ni\~{n}o data and to numerical solutions to the time-delay equation with and without seasonality subtracted and with and without stochastic forcing. We also apply a superposed-epoch analysis to these data to highlight the characteristic waveforms of ENSO episodes as manifested in the Ni\~{n}o signal. The implications of our findings are also discussed in this section and summarized in section (6).

\section{A time-delay equation for ENSO}

An El-Ni\~{n}o episode is characterized by an increase in the sea-surface temperature (SST) in the eastern Pacific, with a strong impact at the coast of Peru. 
Generally, the SST in this area increases during the winter period, but occasionally (often every three to seven years) the temperature increase is more pronounced and this phase is called El-Ni\~{n}o. The opposite phase, a strong decrease in the SST,  is coined  La Ni\~{n}a.
The intensity of El-Ni\~{n}o is higher than the intensity of La-Ni\~{n}a, and the SST distribution is  positively skewed \citep{DK2005}. 
The increase in SST is followed by stronger precipitation, which is distributed over the entire 
Pacific basin, while during La Ni\~{n}a and normal conditions, the precipitation is heavier in the western Pacific. This is also due to the Southern Oscillation (SO) in the atmosphere, which is characterized by lower pressure over the western Pacific during La Ni\~{n}a and normal conditions. According to the Bjerknes hypothesis  \citep{N2011}, there is an initial warming in the SST in the eastern Pacific, which weakens the trade winds from the east and gives rise to westerly wind anomalies, which further increase SST. This positive feedback (which is the El-Ni\~{n}o phase of ENSO) is also accompanied by the formation of the eastward Kelvin wave. 
Concurrently, a westward Rossby wave is formed in the middle of the ocean basin, and after it is reflected at the western boundary of the ocean basin, an eastward  Kelvin wave is formed, which has a cooling effect. This wave  causes the  La Ni\~{n}a phase of ENSO.

A large body of literature is concerned with the dynamical modeling of ENSO (a review can be found in \cite{Dijkstra}). An interesting class of models is described as delay-differential equations, which are linear and autonomous \citep{BH89}, nonlinear, autonomous \citep{SS88}, or nonlinear, periodically forced equations \citep{T94,MCZ}. The latter will be our focus in this paper and can be formulated as an time-delay equation for
the thermocline depth $h(t)$:
\begin{eqnarray}
\frac{dh(t)}{dt}&=& a\tanh (kh\lbrace t-\frac{L}{2c_{K}}\rbrace) \\ \nonumber
&-&
d\tanh (kh\lbrace t-\lbrack\frac{L}{c_{K}}+ 
\frac{L}{2c_{R}}\rbrack \rbrace) 
+c\cos(\Omega t), \label{eq1}
\end{eqnarray}
where $k$ is an ocean-atmosphere coupling parameter, $c_{K}$ is the velocity of the wind-forced Kelvin mode, $L$ is the ocean basin width, and $c_{R}$ is the velocity of the Rossby wave. $\Omega$ is the frequency of the seasonal cycle, while  $a$, $d$ and $c$ are constants.
The  cosine function in equation (\ref{eq1}) accounts for the annual periodicity in the SST data.
This delay differential equation has two time delays: $\tau_{1}=L/c_{K}+L/2c_{R}$, and  $\tau_{2}=L/2c_{K}$. Here, $\tau_{1}$ is the summation over a time it takes a Rossby wave to travel from the middle of the ocean basin
to the western boundary and then be reflected as a Kelvin wave, while $\tau_{2}$ is the transit time for the Kelvin wave which travels from the middle of the basin to reach the eastern Pacific. 

\section{Phase space reconstruction}
Before we describe the test for determinism (or more precisely, for  low-dimensional deterministic dynamics), we briefly explain how the phase space can be reconstructed from scalar time series of  length $N$ by time-delay embedding.
Suppose that the phenomenon under study can be described by a state vector ${\bf z}(t)$ in a phase space of dimension $p$, i.e., $\bf z$ evolves according to an autonomous system of 1st  order ordinary differential equations:
\begin{equation}
 \frac{d{\bf z}}{dt}={\bf f}({\bf z}),\, \, \, {\bf f}: {\cal R}^p\rightarrow {\cal R}^p \label{(eq3)},
 \end{equation}
and that an observed time series $x(t)$ is generated by the measurement function $g: {\cal R}^p\rightarrow {\cal R}$:
\begin{equation}
x(t)=g({\bf z}(t)). \label{(eq4)}
\end{equation}
Further, assume that the dynamics takes place on an invariant set (an attractor) ${\cal A} \subseteq {\cal R}^p$ in phase space, and that this set has box-counting fractal  dimension $d$. Since the dynamical system  uniquely defines the entire phase-space trajectory once the state ${\bf z}(t)$ at a particular  time $t$ is given, we can define uniquely an $m$-dimensional measurement function, 

\begin{equation}
{\bf g}: {\cal A}\rightarrow {\cal R}^m, \, \, {\bf g}({\bf z})=(x(t),x(t+\tau),\ldots,x(t+(m-1)\tau)), \label{(eq5)}
\end{equation}
where  the vector components are given by equation (\ref{(eq4)}), and $\tau$ is a  time delay of our choice. If the invariant set $\cal A$ is compact (closed and bounded),  $g$ is a smooth function and $m>2d$, the map given by equation (\ref{(eq5)}) is a topological embedding (a one-to-one  continuous map) between $\cal A$ and ${\cal R}^m$. The condition $m>2d$ can be thought of as a condition for the image ${\bf g}({\cal A})$ not to intersect itself, i.e. to avoid that two different states on the attractor $\cal A$ are mapped to the same point in the $m$-dimensional embedding space ${\cal R}^m$. If such an  embedding is  achieved, the trajectory ${\bf x}(t)={\bf g}({\bf z})$ (where ${\bf g}({\bf z})$ is given by equation (\ref{(eq5)})) in the embedding space is a complete mathematical representation of the dynamics on the attractor. Note that the dimension $p$ of the original phase space is irrelevant for the reconstruction of the embedding space. The important thing is the dimension $d$ of the invariant set $\cal A$ on which the dynamics unfolds.

\section{Test for determinism}
\begin{figure}
\begin{center}
\includegraphics[width=6cm]{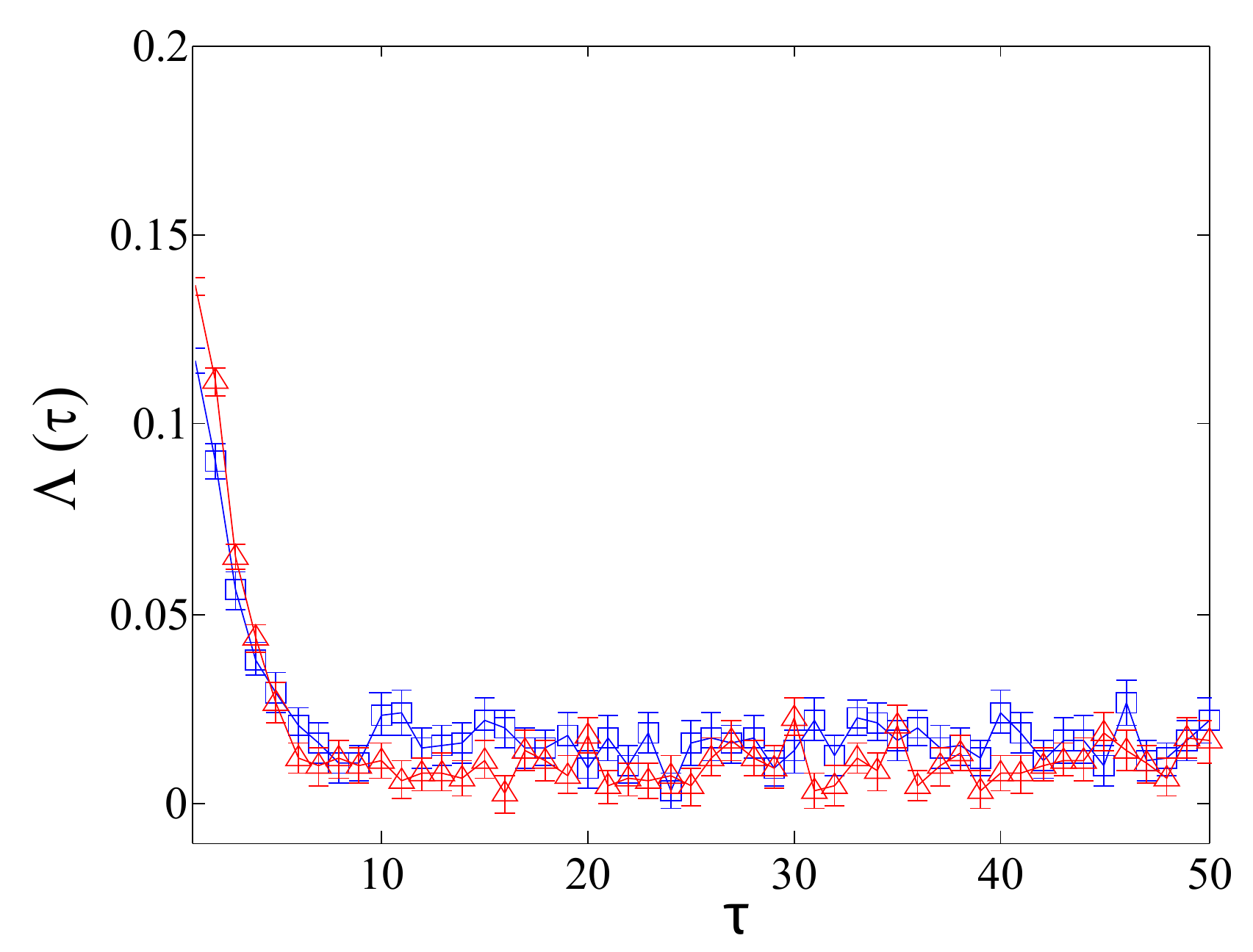}
\caption{$\Lambda (\tau)$ averaged over results computed from an ensemble of  ten realization of the  O-U process (squares), and  for the realizations with randomized phases (triangles). Error bars denote standard errors of the {\em mean} $\Lambda(\tau)$ (which is $(1/\sqrt{10})\times $the standard deviation of the distribution of the ten samples).}
\end{center}
\end{figure}
This method was recently successfully applied to the studies of magnetospheric organization during magnetospheric storms and substorms \citep{ZR1,ZR2}.
When a system is low-dimensional deterministic, the direction of the trajectory (its tangent) is a function of the position in the reconstructed phase space (from equation \ref{(eq3)}). This means that  trajectories emanating from points in a small neighborhood in phase space have almost parallel directions. On the other hand, corresponding trajectories in a stochastic or high-dimensional system have directions in a low-dimensional embedding space which are not uniquely dependent on the position in this space, and therefore the tangent can  have a  different direction the next time it recurs to the same neighborhood. Let $b$ denote a small time increment and envisage a portion of phase space spanning the entire attractor divided into an enumerable set of  small ``boxes" of size corresponding to the length of the trajectory increment:

\begin{eqnarray}
\Delta {\bf x}(t)&=&[ x(t+b)-x(t), x(t+\tau+b)-x(t+\tau),\ldots , \nonumber  \\ 
& &x(t+(m-1)\tau+b)-x(t+(m-1)\tau)] , \label{(eq6)}
\end{eqnarray}
\noindent 
The  tangent for the $k$'th pass of the trajectory through box $j$ is the unit vector ${\bf u}_{k,j}=\Delta {\bf x}_{k,j}(t)/\vert \Delta {\bf x}_{k,j}(t)\vert$.
The  estimated averaged displacement  vector in the box is

 \begin{equation}
{\bf V}_{j}=\frac{1}{n_{j}}\sum_{k=1}^{n_{j}}{\bf u}_{k,j}, \label{(eq7)}
\end{equation}
where $n_j$ is the number of passes of the trajectory through box $j$.  If the embedding dimension is sufficiently high and in the limit of vanishingly small box size, the trajectory directions should be aligned and the length  $V_j\equiv \vert  {\bf V}_{j}\vert=1$. In the case of deterministic dynamics and finite box size, ${\bf V}_{j}$ will not depend very much on the  number of passes $n_j$, and $V_j$ will  converge to $1$  as $n_j\rightarrow \infty$. In contrast, for the trajectory of a random process, where the direction of the next step is completely independent of the past, $V_j$ will decrease with $n_j$ as $V_j\sim n_j^{-1/2}$.
The  degree of determinism of the dynamics can be assessed by exploring the dependence of $ V_j$ on $n_j$. In practice, this can be done by computation of  the averaged displacement vector length:

 \begin{equation}
 L_n\equiv\langle  V_{j}  \rangle_{n_j=n}, \label{(eq8)}
 \end{equation} where the average is done over all boxes with same number $n$ of trajectory passes. 

\begin{figure}
\begin{center}
\includegraphics[width=6cm]{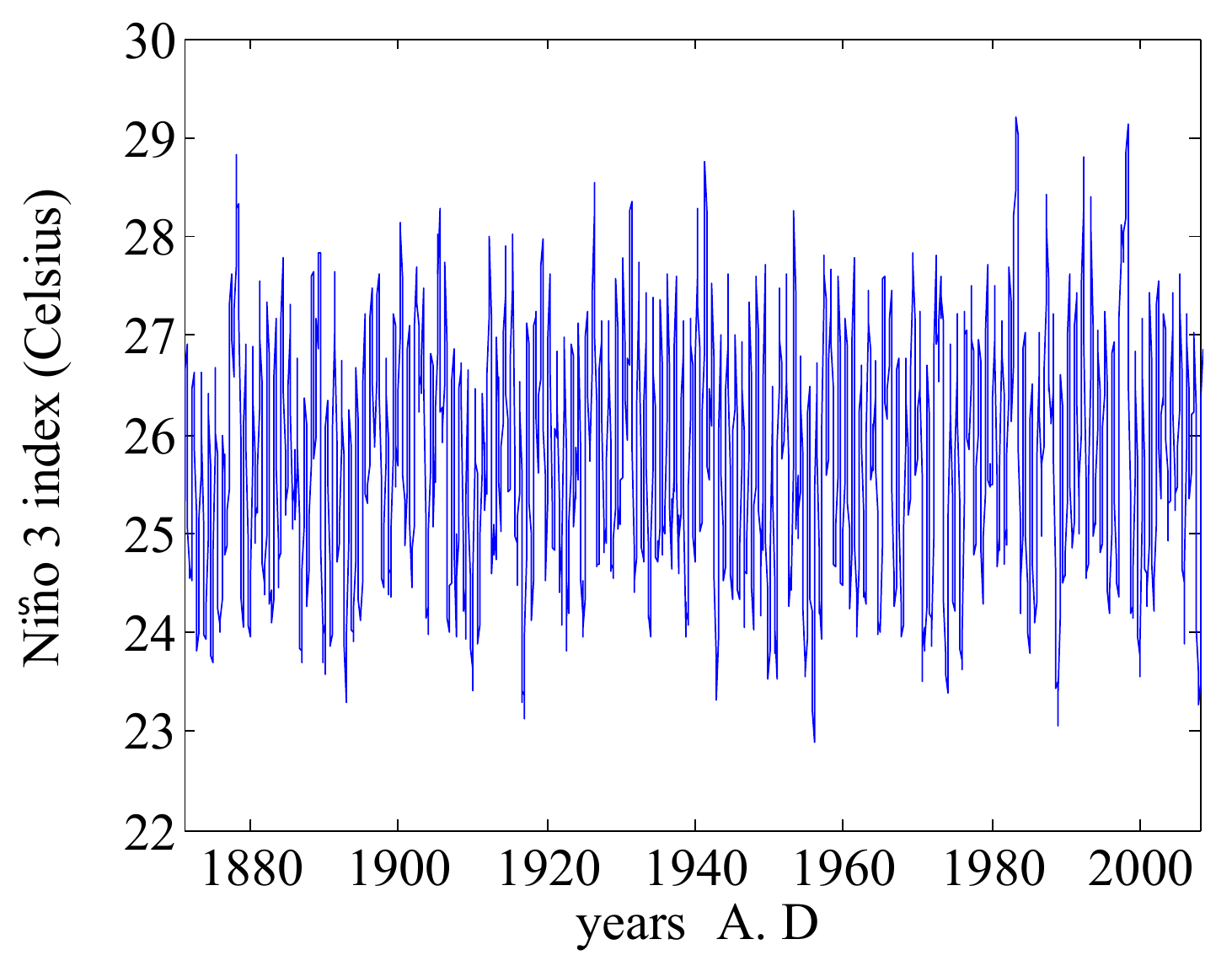}
\caption{The Ni\~{n}o 3 monthly time series.}
\end{center}
\end{figure}
Let us recall that we are  describing a test to distinguish signals described by low-dimensional dynamical systems, i.e., signals that in the continuous-time limit are solutions to differential equations and therefore continuous, and signals described as a stochastic process. The archetype of a random and continuous stochastic process is the Wiener process (Brownian motion), which in discrete time is a random walk. The random walk has random increments, and hence the displacement vectors $\Delta {\bf x}(t)$ in an $m$-dimensional embedding space will have random directions. Hence, for a random walk $L_{n\rightarrow \infty}=0$. For finite $n$, however, there will be a finite statistical spread of  $V_j$, and 
as shown in \cite{KG93}, the average displacement of $n$ passes in {\it m}-dimensional phase space is \begin{equation}
L_n=R_n\equiv \frac{1}{\sqrt{n}}(\frac{2}{m})^{1/2}\frac{\Gamma\lbrack (m+1)/2\rbrack}{\Gamma(m/2)},\label{(eq9)}
\end{equation} 
where $\Gamma$ is the  gamma function. 
The deviation in $\langle V_{j}\rangle$  between a given time series and a random-walk can be characterized by a single number given by the weighted average over all boxes of the quantity,
\begin{equation}
\Lambda(\tau)\equiv \frac{1}{\sum_{j} n_{j}}\sum_{j} n_{j} \frac{\langle V_j\rangle ^{2}(\tau)-R_{n_j}^{2} }{1-R_{n_j}^{2}},\label{(eq10)}
\end{equation}
where we have explicitly highlighted that the averaged displacement $\langle V_j\rangle(\tau)$ of the trajectory in the reconstructed phase space depends on the time delay $\tau$.   For a completely deterministic signal we have $\Lambda(\tau)=1$, and for a completely random signal $\Lambda(\tau) =0$, hence this quantity can be  considered as a measure of determinism. 

In Figure 1, we show  $\Lambda(\tau)$ averaged over ten numerical realizations of the Ornstein-Uhlenbeck (O-U) stochastic process, for embedding dimension is $m=8$, and $b=1$. The O-U process is described  by the stochastic equation:
\begin{equation}
dS_{t}=-\lambda S_{t}+\sigma dW_{t},
\end{equation}
where $W(t)$ is the Wiener process. It is a more physically realistic random process for many phenomena than the Wiener process, since the damping term $-\lambda S_{t}$ makes it bounded. In Figure 1 we also show mean $\Lambda(\tau)$ computed from the same ten realizations, but after randomization of the phases of the Fourier coefficients. This randomization leaves the power spectral density, and hence the autocorrelation function, unchanged. Hence, $\Lambda(\tau)$ should also be unchanged for a random process, which Figure 1 demonstrates. 
On the other hand, for a signal from a low-dimensional chaotic system, which has to be nonlinear to be chaotic, the randomization of phases will destroy the nonlinear coupling between Fourier modes and make $\Lambda (\tau)$ more similar to a random signal, i.e., it will be reduced compared to  $\Lambda (\tau)$ for the original signal. Examples of this were shown in \cite{KG92,KG93}. 

\section{Results}

We analyze Ni\~{n}o 3 data (see Figure 2) obtained from the sea ice and SST data set (HadISST1) \citep{R2003}. Ni\~{n}o 3 data is the area-averaged SST from 5S-5N and 150W-90W, with monthly resolution and the time span between 1871 and 2008.
We compute $\Lambda$ as a function of time-delay, which is also used in the phase-space reconstruction procedure. The embedding dimension is $m=10$, $b=1$, i.e., the box size is equal to the average distance between two successive points on the reconstructed phase-space trajectory. It has been shown in \cite{KG93} that increasing embedding dimension can increase $\Lambda$ in deterministic systems, while it should not influence $\Lambda$ in stochastic systems. In principle, higher $m$ is better, but we also have to consider the number of data points available for the test, and this decreases with increasing $m$ (from equation \ref{(eq5)}). In our analysis, $m=10$ seems a suitable choice of embedding dimension.
In Figure 3 we observe that  $\Lambda(\tau)$ exhibits spiky dips at those $\tau$ where the autocorrelation function $r(\tau)$ has an extremum. This is a spurious   feature of the technique which is explained in \cite{KG93}. We also plot  $\Lambda (\tau)$ averaged over  an ensemble of ten surrogate time series with randomized phases. 

From Figure 3, we can conclude that the dynamics underlying the Ni\~{n}o 3 data  is dominated by a nonlinear, low-dimensional component, since $\Lambda (\tau)$ for the phase-randomized time series is considerably reduced compared to  that computed for the original Ni\~{n}o 3 data. 
However, as we will demonstrate next, the nonlinear and low-dimensional behavior is rather  trivial and derives from the seasonal cycle. This cycle can be represented by the climatology, which is 
the mean over all data for every month of the year. In Figure 4 we plot the Ni\~{n}o 3 climatology along with the  sinusoidal climatology used in equation (\ref{eq1}).
Next, we subtract the climatology from the Ni\~{n}o 3 data, and compute $\Lambda(\tau)$ again (Figure 5 (a)).  $\Lambda(\tau)$ is strongly reduced and randomization of phases makes no discernible difference, indicating the the positive test for determinism in the original Ni\~{n}o 3 time series is an effect of the seasonal cycle. 
 \begin{figure}
\begin{center}
\includegraphics[width=6cm]{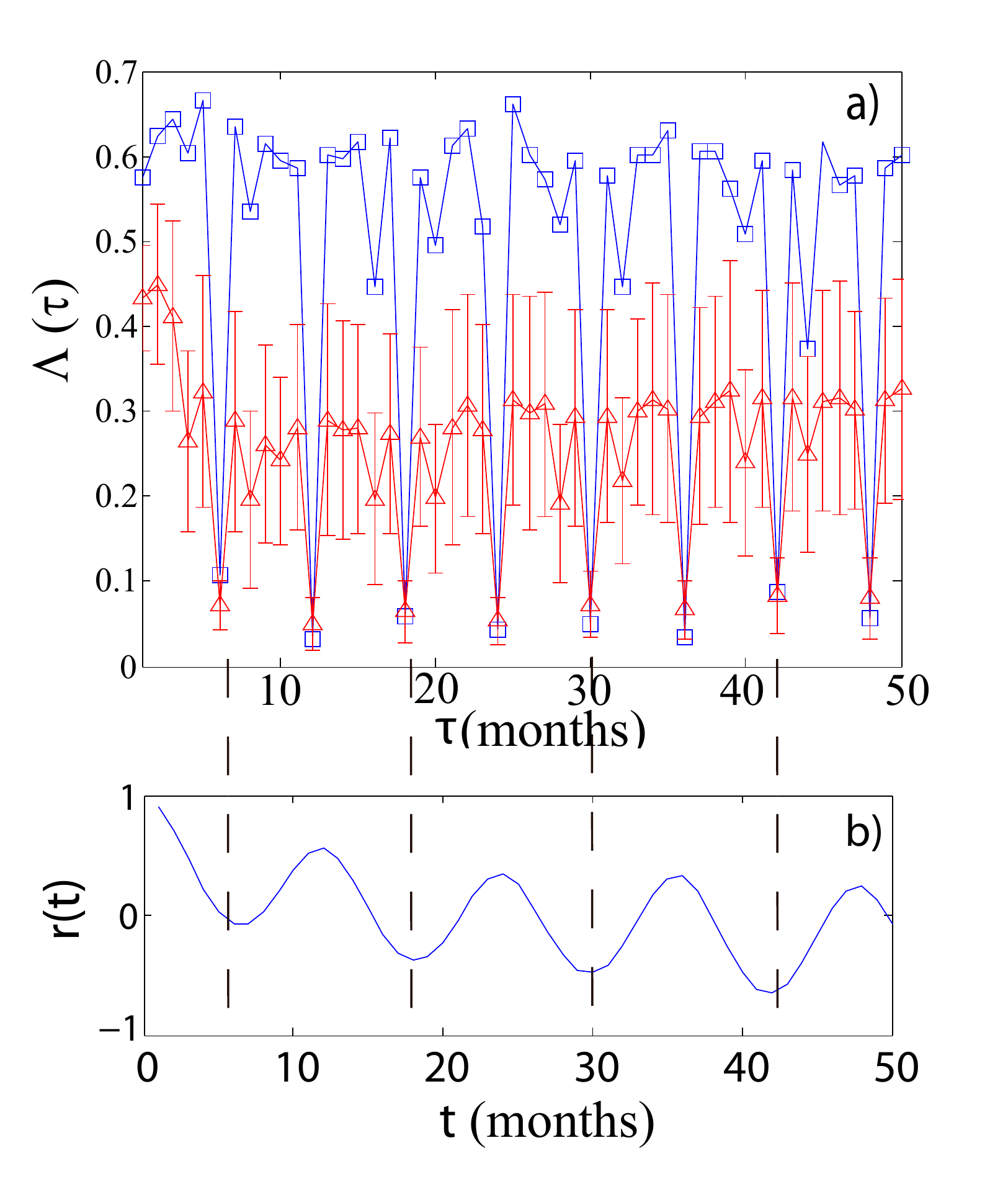}
\caption{ (a): $\Lambda(\tau)$ for the Ni\~{n}o 3 time series (squares) and $\Lambda(\tau)$ averaged over ten different phase-randomized versions of this time series (triangles). The error bars of the latter denotes the standard deviation of $\Lambda(\tau)$ over the distribution of these ten samples. (b): autocorrelation function $r(\tau)$  for the Ni\~{n}o 3 time series.}
\end{center}
\end{figure}

\begin{figure}
\begin{center}
\includegraphics[width=6cm]{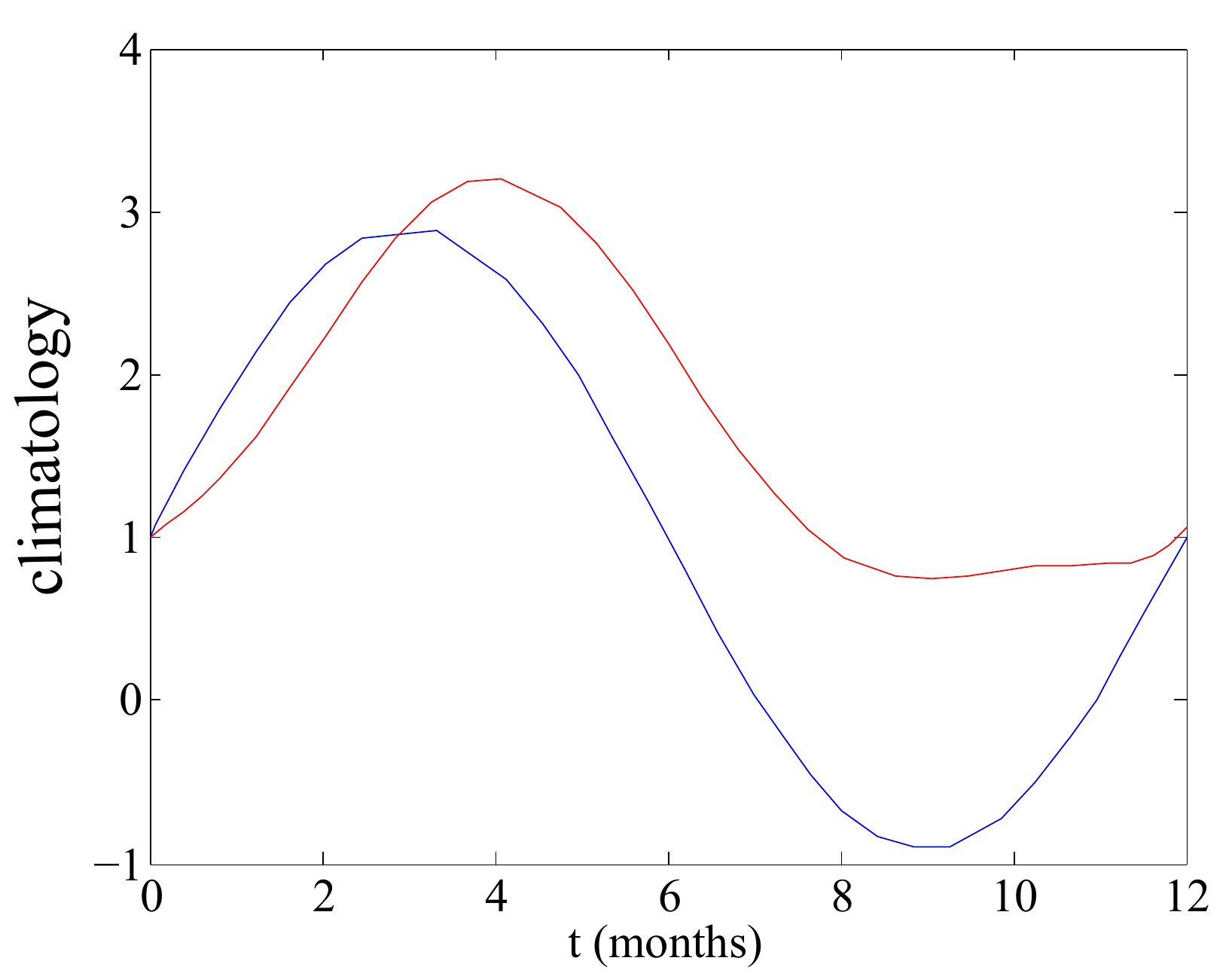}
\caption{Climatology for Ni\~{n}o 3 time series (red) and the sinusoidal climatology used in equation (\ref{eq1}) (blue).}
\end{center}
\end{figure}

\begin{figure}
\begin{center}
\includegraphics[width=6cm]{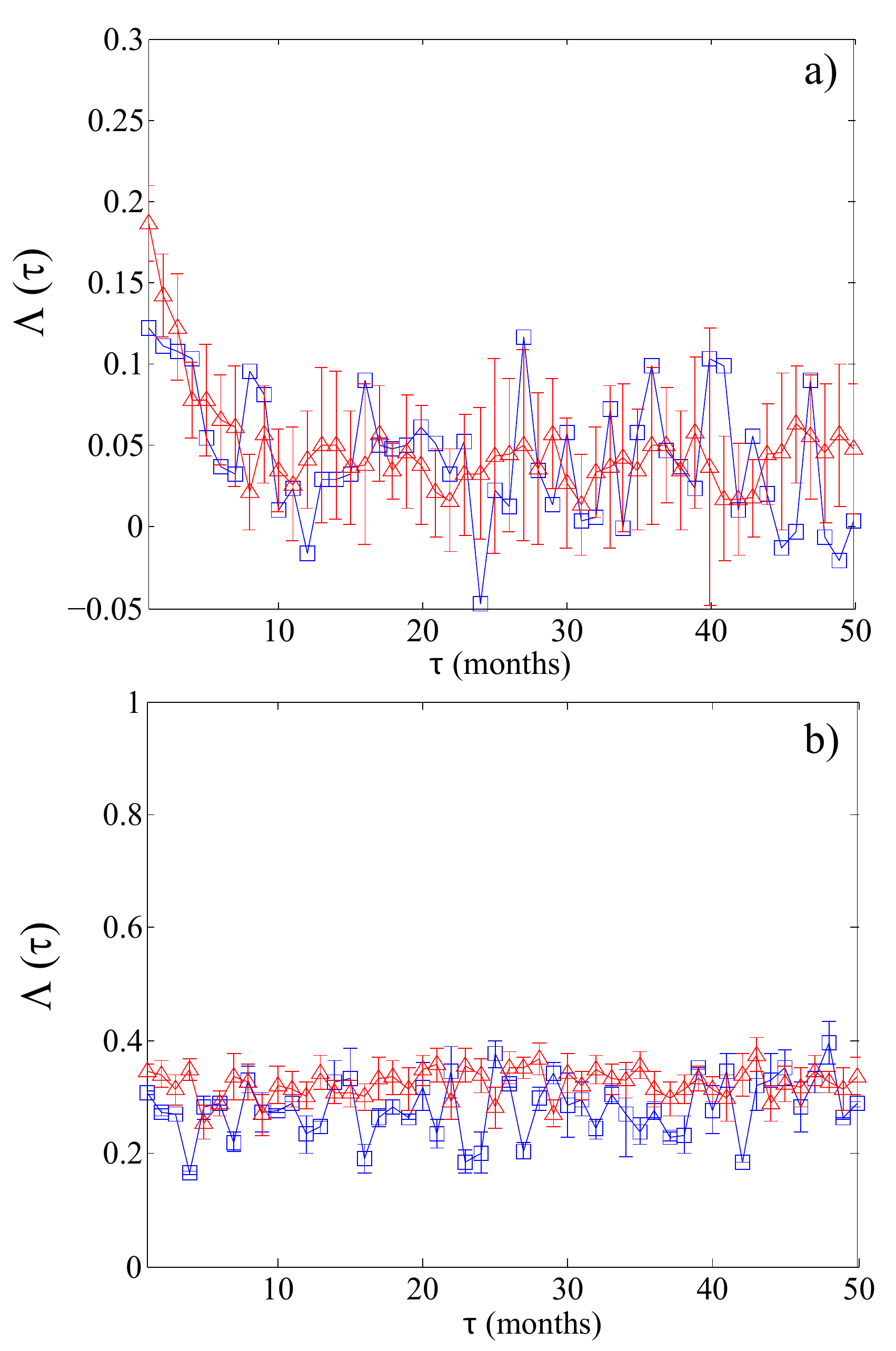}
\caption{(a): $\Lambda(\tau)$ for original Ni\~{n}o 3 time-series   with climatology subtracted (blue squares), and $\Lambda(\tau)$ averaged over ten realizations of phase-randomized versions of this time series (red triangles). The error bars of the latter denotes the standard deviation of $\Lambda(\tau)$ over the distribution of these ten samples.  (b): $\Lambda(\tau)$ averaged over ten realizations of time series generated by equation (\ref{eq1}) with a white-noise stochastic forcing and with climatology subtracted (blue), and the same for ten phase-randomized versions (red). In (b) the error bars denote the standard deviation of the {\em mean} $\Lambda(\tau)$.}
\end{center}
\end{figure}

\begin{figure}
\begin{center}
\includegraphics[width=6cm]{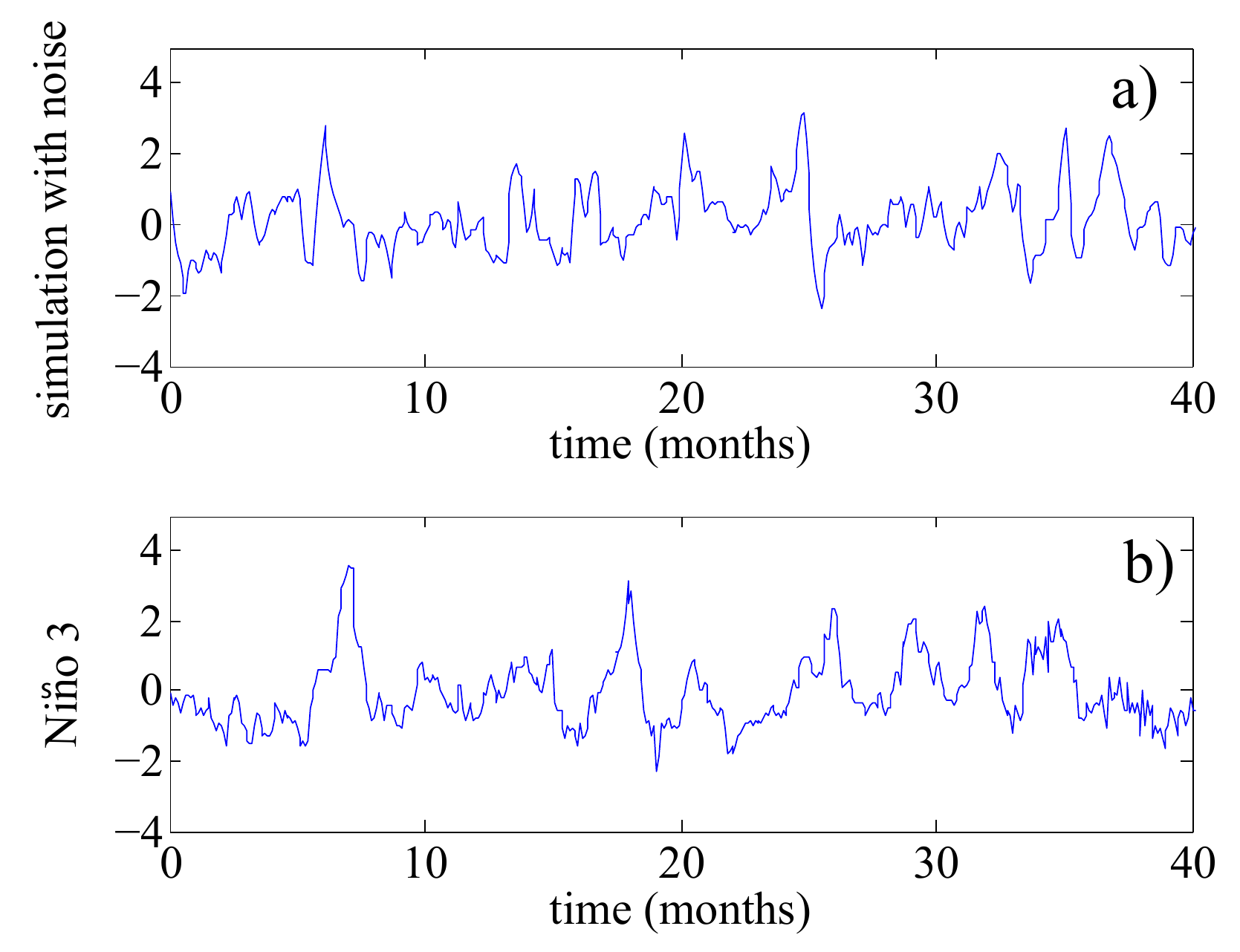}
\caption{Time series analyzed in Figure 5.: (a): simulation. (b): Ni\~{n}o 3.}
\end{center}
\end{figure}

\begin{figure}
\begin{center}
\includegraphics[width=6cm]{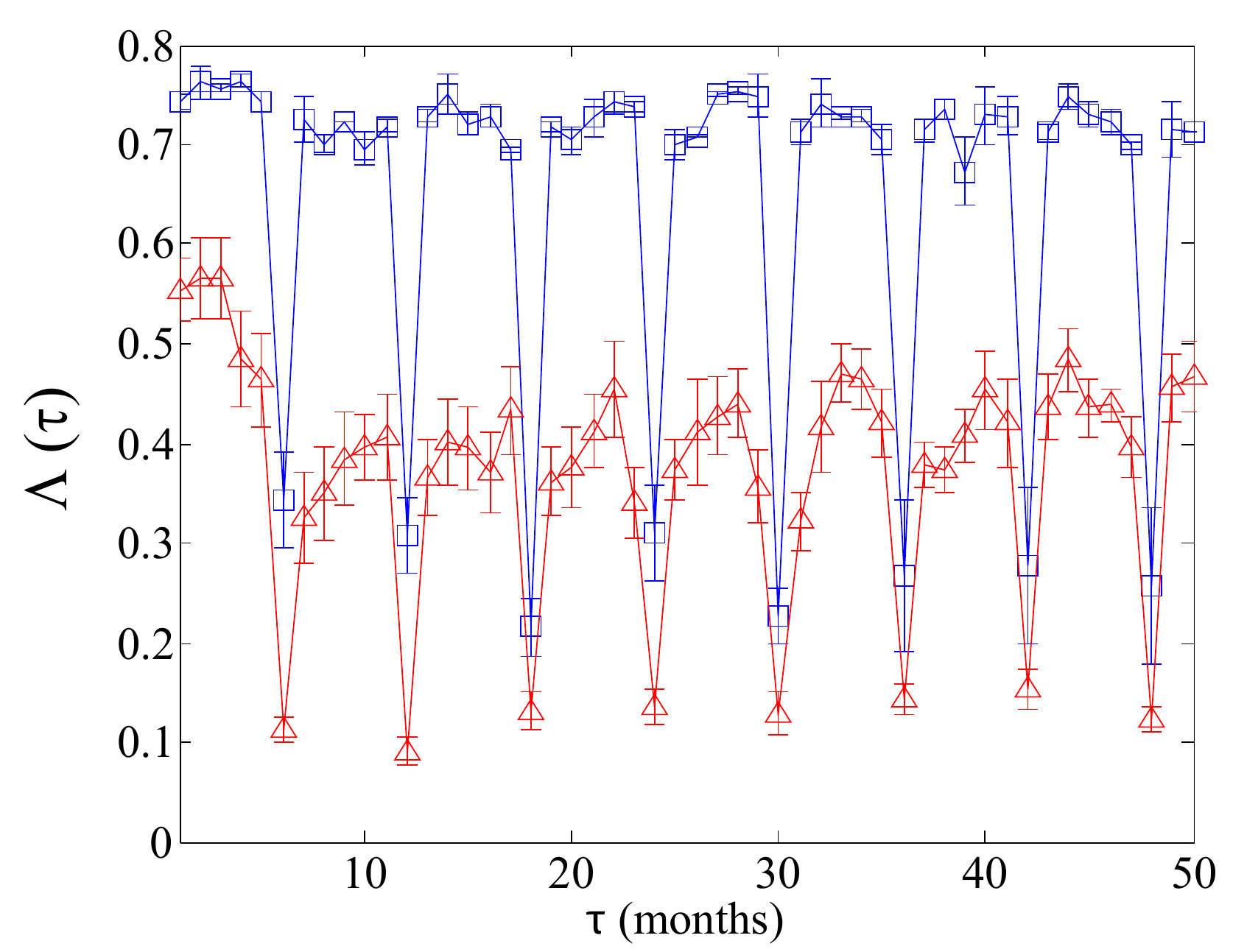}
\caption{$\Lambda(\tau)$ averaged over results computed from ten simulated time series like that shown in Figure 6 (a) with climatology included (blue squares). After phase-randomization (red triangles). The error bars denote the standard deviation of the {\em mean} $\Lambda(\tau)$.}
\end{center}
\end{figure}
 
 \begin{figure}
\begin{center}
\includegraphics[width=6cm]{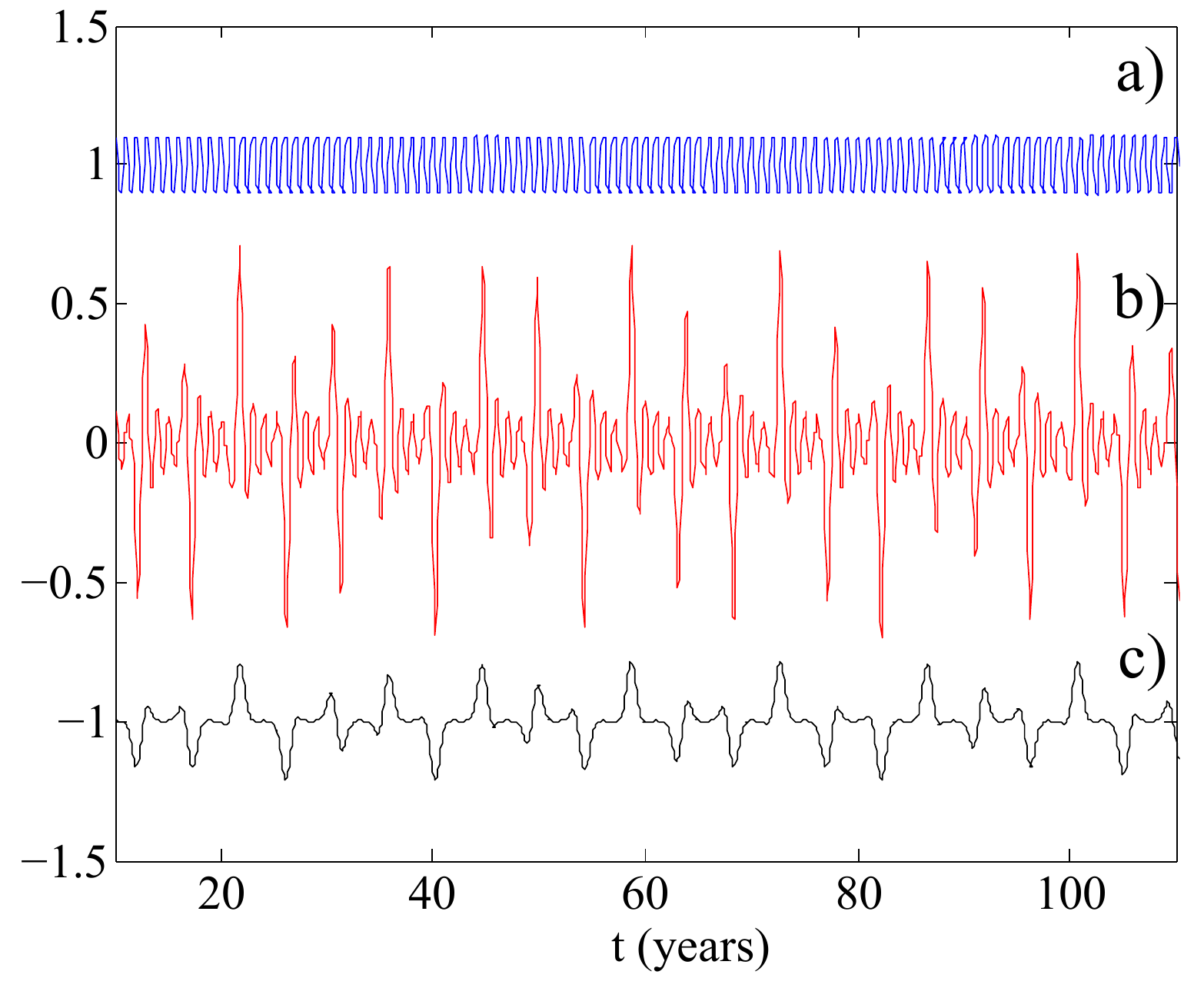}
\caption{Simulation of time-delay equation with no stochastic forcing. (a):    $\tau_{1}=0.4088$. (b):  $\tau_{1}=0.4198$. (c): the signal in (b) de-seasonalized by wavelet filtering.}
\end{center}
\end{figure}

\begin{figure}
\begin{center}
\includegraphics[width=6cm]{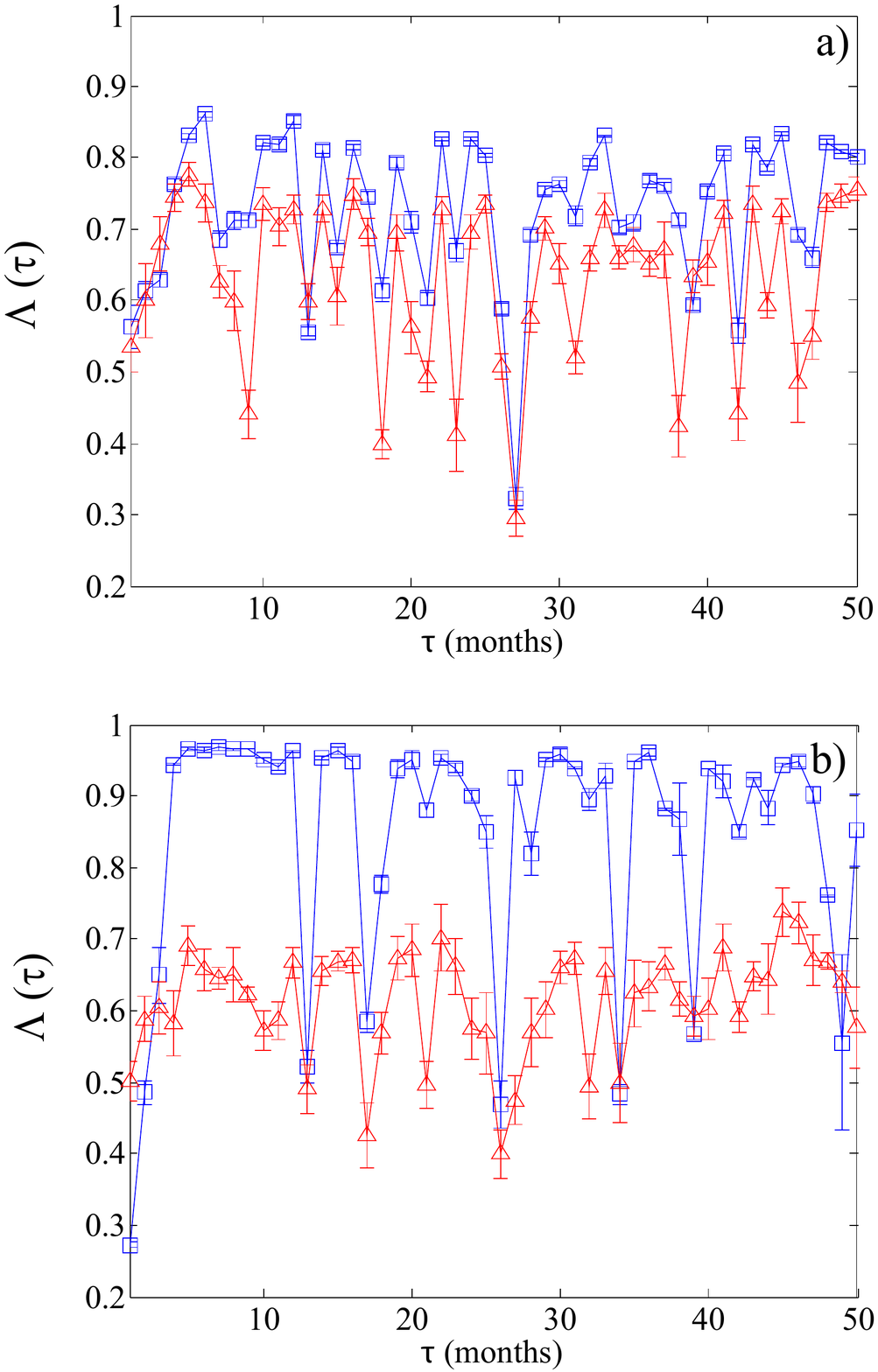}
\caption{ (a): $\Lambda(\tau)$ averaged over results computed from ten simulated time series like that shown in Figure 8(b) (blue squares). After phase-randomization (red triangles). (b): $\Lambda(\tau)$ averaged over results computed from ten simulated time series like that shown in Figure 8(c) (blue squares). After phase-randomization (red triangles). Error bars denote standard deviation of the ensemble mean.}
\end{center}
\end{figure}

\begin{figure}
\begin{center}
\includegraphics[width=6cm]{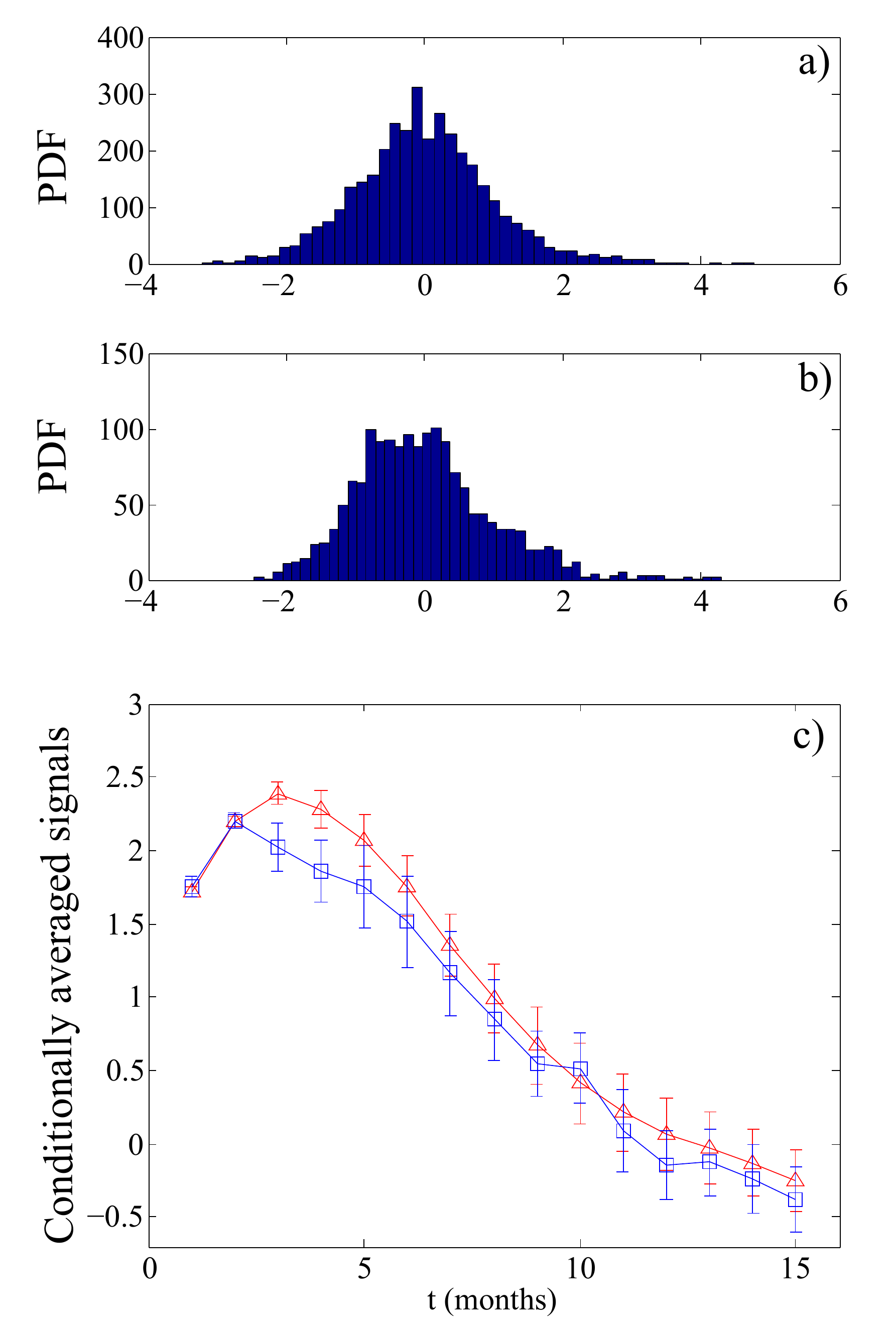}
\caption{(a): Histogram of signal in Figure 6(a). (b): Histogram of signal in Figure 6(b). Conditionally averaged evolution of ENSO signal after onset of ENSO episode. For simulation with stochastic forcing shown in Figure 6(a) (red triangles). For Ni\~{n}o 3 index shown in Figure 6(b) (blue squares). The error bars denote the standard deviation of the $\langle s_n(\delta t)\rangle$.}
\end{center}
\end{figure}
This implies that the Ni\~{n}o 3 time series with the seasonal cycle subtracted can be described a stochastic process similar to the O-U process for which results were presented in Figure 1. It does not imply, however,  that the time-delay equation (\ref{eq1}) has to be abandoned, since inclusion of  stochastic forcing term
may give rise to such stochastic dynamics and produce results compatible with the test for determinism in Ni\~{n}o 3 data. In the following we shall first show  that such a time-delay equation with stochastic forcing is able to produce time series with these properties. Next, we shall demonstrate that in absence of stochastic forcing, the time-delay model tends to produce low-dimensional, deterministic dynamics, even when seasonality is removed.

In order to investigate the effect of such stochastic forcing we add a  term $\sigma w(t)$, where $w(t)$ is a Gaussian white noise with unit variance, to equation (\ref{eq1}) and solve the equation numerically with the parameters $a=0$, $d=1, k=100, c=1.3, \sigma=0.7$ and $\tau_{1}=0.2$. Figure 5(b) shows $\Lambda(\tau)$ ensemble-averaged over ten realizations of the simulation after the  climatology has been subtracted. In the same plot we show  $\Lambda(\tau)$ averaged over the ten corresponding phase-randomized realizations. The low values of $\Lambda(\tau)$, and their non-responsiveness to phase randomization, indicate that for these parameters the de-seasonalized solutions to the stochastically  forced equation (\ref{eq1}) have the character of a stochastic process  similar to the de-seasonalized Ni\~{n}o 3 time series.
A qualitative similarity is also apparent from  Figure 6, where  we plot a realization of this process along with the  Ni\~{n}o 3 time  series, both with the climatology subtracted.  Both time series have been normalized to have unit variance.
This similarity is also apparent when $\Lambda(\tau)$ is computed from the  simulation without subtracting the seasonal cycle, as shown  by comparing Figure 7 with Figure 3. 
The model described above, but without stochastic forcing ($\sigma=0$), was studied by \cite{G2008}. They used the same choice of parameters $a$, $c$ and $d$ as above, but varied the parameters $k$ and $\tau_1$. They computed and  plotted the global maximum of the thermocline depth $h(t)$ as a function of $\tau_{1}$ and  $k$, and observed discontinuities in this parameter space separating regions of  regular solutions strictly governed by the seasonal forcing and more ENSO-like irregular solutions, indicating a structural instability of the model. In Figure 8 (a)  we have plotted such a regular solution obtained for $\tau_{1}=0.4088$, and in  Figure 8 (b) an irregular one on the other  side of the parameter-space  discontinuity; for $\tau_{1}=0. 4198$. In both of these cases, $c=1$,  $k=100$ and $d=1$. This irregular solution exhibits a seasonal cycle of variable strength interrupted by stronger episodical oscillations with some reminiscence of observed ENSO episodes. However, by inspection of the time series the similarity to the Ni\~{n}o 3 signal is not too convincing. As can be seen from Figure 6, the solution with stochastic forcing added exhibits more qualitative similarity to the Ni\~{n}o 3 signal than the signal in Figure 8(b). This qualitative discrepancy is also reflected by the test of determinism.  Figure 9(a) shows $\Lambda(\tau)$ for the signal in Figure 8(b) and for its phase-randomized version. The determinism is higher than the one for the Ni\~{n}o 3 signal shown in Figure 3(a), and it is less reduced by phase randomization. The former is obviously due to the higher degree of randomness apparent in the Ni\~{n}o 3 signal. The lack of reduction of determinism after phase-randomization  of the Figure 8(b)-signal is due to the strong seasonal oscillation. Since a major fraction of the power resides in the annual Fourier component the phase randomization will change the phase of this component, but the signal will still have a deterministic appearance after this change. This is all quite trivial, so the crucial test is to compute determinism after de-seasonalization. For the signal in Figure 8(b) subtraction of the climatology does not offer an effective filtering of the seasonal component, since its amplitude varies a lot. For the same reason Fourier filtering is also ineffective. In Figure 8(c) we show the de-seasonalized signal after application of a mexican-hat wavelet filter. This signal represents the ``true" ENSO episodes according to the time-delay model without stochastic forcing. The computed determinism $\Lambda(\tau)$ of this signal is shown in Figure 9(b). The determinism is very high, and slows clearly that the ENSO episodes in this model are the result of deterministic, low-dimensional dynamics. The strong reduction of determinism after phase-randomization demonstrates that this dynamics is nonlinear, and (since the timing of the episodes seems random) chaotic. This is in strong contrast to the results shown in Figure 5 for de-seasonalized  Ni\~{n}o 3 and  signals generated by the time-delay model with stochastic forcing, which show signals dominated by a stochastic component.

In the following, we go back to the time-delay simulation with stochastic forcing and demonstrate in two more examples its similarity to the Ni\~{n}o 3 data.
In Figure 10(a)  we plot a histogram for the Ni\~{n}o 3 data and for the simulation after the climatology has been subtracted (the signals in Figure 6). The ENSO events contribute to the tails of these distributions, which are somewhat heavier on the positive side due to the relative strength of El Ni\~{n}o compared to La Ni\~{n}a. The average time-evolution of ENSO events in Ni\~{n}o 3 and simulation can be investigated by means of a superposed-epoch analysis (conditional averaging). This analysis works as follows: Consider a signal $s(t)$. We define the onsets of ENSO events as the times $t_n$, $n=1,\ldots,N$, for which the signal ascends through two standard deviations from the mean. Then we produce an ensemble of conditional signals $s_n(\delta t)=s(t_n+\delta t)$ and produce the conditional average $\langle s_n(\delta t)\rangle \equiv (1/N)\sum_{n=1}^{N}s_n(\delta t)$. Figure 10(c) displays the average ENSO structure for Ni\~{n}o 3 (squares) and simulation  (triangles) computed this way, and shows that the simulation predicts fairly well the average evolution of ENSO up to at least  15 months after the onset of the episode.

\section{Conclusion}
From  of a model for ENSO activity similar to the one studied here,  \cite{ET95} came to the conclusion that depending on the strength of the coupling between the ocean and the atmosphere, the dynamics could undergo quasi-periodicity routes to chaos.
The same authors suggested that the ENSO might be described as a low-dimensional chaotic dynamics driven by the seasonal cycle, where the appearance of the chaos is due to the nonlinear resonance between the natural oscillator of the atmosphere-ocean coupling and the seasonal cycle. Further, the irregular jumps of the state between different resonances should be the cause of the chaos. Unfortunately, SST time series in the eastern Pacific are too short for chaos to be proven by standard methods.
 
In this paper we have applied a simple test for determinism to show that Ni\~{n}o 3 data, which is the proxy of SST in the eastern Pacific, is most adequately described as a stochastic process after the seasonal cycle has been removed, implying that ENSO is a consequence of mainly stochastic and not low-dimensional, chaotic dynamics. 
A similar conclusion was made by \cite{BW2002}, where a test for determinism was applied to the Southern Oscillation index (SOI) series. SOI measures a pressure difference between Tahiti and Darwin in the Pacific Ocean and can also be used as a proxy for the El-Ni\~{n}o dynamics. 

Despite the stochastic nature of the Ni\~{n}o 3 signal, we have demonstrated here that low-dimensional, deterministic dynamics may also be involved. The equatorial-wave equation from \cite{T94} exhibits such dynamics without stochastic forcing, but  by adding stochastic forcing to this equation and comparing determinism and  average ENSO structure, with those of Ni\~{n}o 3 data, we conclude that this statistical-dynamical model can reproduce important aspects of ENSO dynamics.

\end{article}
\end{document}